AUTHOR-ACCEPTED MANUSCRIPT

# Analyzing Simultaneous Charging Factors of Electric Vehicles Using Monte Carlo Simulation: A Case Study in Phu Quoc City, Vietnam

Tien-Khai Nguyen · Trong Nghia Le

# Analyzing Simultaneous Charging Factors of Electric Vehicles Using Monte Carlo Simulation: A Case Study in Phu Quoc City, Vietnam

Tien Khai Nguyen[1,*], and Trong Nghia Le[1]

ABSTRACT

This study estimates the simultaneous charging factor (SCF) of electric vehicles (EVs) in Phu Quoc City, Vietnam, using a Monte Carlo simulation. Three planning scenarios (2030, 2035, and 2040) are developed from population-based EV uptake assumptions and vehicle-specific charging probability profiles over 24 hours, reflecting distinct operating patterns of motorbikes, cars, taxis, and buses. For each vehicle class and scenario, 1,000 simulation runs are performed to compute SCF and its uncertainty, reported via 95% confidence intervals. Results indicate that SCF decreases as EV penetration increases, implying more temporally dispersed charging demand and a reduced likelihood of extreme coincidence. For electric buses, SCF declines from 0.76 (2030) to 0.59 (2040), and other vehicle groups show similarly decreasing trends across the horizon. These SCF estimates provide practical inputs for grid-aware charging infrastructure planning, including diversity factors for sizing distribution transformers and feeders, identifying peak-hour stress, and informing demand management measures such as time of use tariffs and basic smart charging controls.

## 1. INTRODUCTION

The transition to electric vehicles (EVs) helps reduce greenhouse gas emissions and pollution, though the need for recharging remains a challenge, and interest in EVs is growing due to high carbon dioxide emissions from internal combustion engine vehicles (ICEVs) [1]. Many countries view EVs as crucial for achieving Net Zero Emissions by 2050 [2]. Vietnam is also accelerating EV deployment. With strong support from VinFast, the country is rapidly expanding its charging network in the region. VinFast plans to install around 150,000 charging ports across all provinces and cities in Vietnam [3]. This exceeds reported figures for nearby countries such as Thailand (8,700 ports) and Singapore (3,600 ports), and is even higher than in countries like the Netherlands (over 120,000), France (about 84,000), and Germany (77,000) [4]. However, as more EVs appear on the roads, the need for charging points increases, which can significantly change load profiles and put additional stress on the distribution grid [5]–[7]. Accordingly, the timing and control of EV charging have become a key issue for both researchers and policymakers [8].

Phu Quoc is a strategic coastal city in southern Vietnam. At the time of data collection, it was administered as Phu Quoc City under Kien Giang Province. Following a national administrative reform in July 2025, it was reorganised as the Phu Quoc Special Administrative Zone under An Giang Province; however, for consistency with the study period, this paper continues to use the term Phu Quoc City. According to Decision No. 150/QD-TTg dated 6 February 2024, the Prime Minister has approved a plan for Phu Quoc's development to 2040. Under this plan, the city is expected to become a major centre for tourism, business and services, playing a key role in the Mekong Delta economy. Phu Quoc is envisioned as a modern, smart city that emphasises environmental protection and long-term growth, with new urban areas, eco-tourism zones, clean industries and tourism-supporting services developed in parallel with the conservation of natural resources and cultural heritage [9]. Building more charging stations in Phu Quoc not only supports current EV demand but also helps create a smarter electricity grid for the future [10]. Even though charging stations are key to encouraging EV use, policies for developing them remain limited. This has slowed the move from petrol vehicles to electric ones [11].

Previous research on EV charging has primarily examined factors that influence charging demand, such as time-of-day patterns and user behavior, and has proposed solutions including time-of-use tariffs, smart charging schemes, and improved the energy management strategies [12]–[14]. Within this literature, a key indicator is the SCF, defined as the proportion of EVs charging at the same time; high SCF values can create significant stress on distribution networks, whereas low values may indicate under-utilisation of charging infrastructure [15]. Despite this, most existing studies still concentrate on aggregate load profile estimation, pricing mechanisms, or feeder hosting capacity, rather than on a detailed characterisation of charging simultaneity across different vehicle classes.

[1]*Department of Electrical and Electronics Engineering, HCMC University of Technology and Engineering, Vietnam.*
**Corresponding author:* Tien Khai Nguyen; Phone: +84-961-432-199; Email: 2430605@student.hcmute.edu.vn.

MCS provides a robust framework for modelling and analysing stochastic, uncertain, and complex systems, including EV charging networks, and it has been widely used to study charging habits and evaluate charging strategies by considering factors like vehicle arrival times, charging durations and user behavior [16]. However, these studies rarely distinguish between multiple vehicle types in a specific urban context [17, 18].

Recent regional studies have used Monte Carlo–based methods to analyse EV charging demand and grid loading in realistic networks. Applications include comparing uncoordinated and coordinated charging in a Thai distribution system, characterising plug-level power use at public stations in an Indonesian urban grid, predicting spatiotemporal charging loads for planning in a coupled transport and power network in China and estimating coincident factors for public e-buses and taxis in Greater Ho Chi Minh City [19]–[22]. These contributions demonstrate that stochastic charging models can capture uncertainties in EV arrival times, state-of-charge, dwell times, and user behavior, and can be embedded in detailed distribution network simulations to quantify impacts on transformer loading, voltage profiles and technical losses.

Despite the growing body of work on EV load profiles and charging management, relatively few studies have quantified simultaneity factors for different vehicle classes using MCS methods in a real urban context. This paper addresses this gap by estimating SCFs for electric motorbikes, cars, taxis, and buses in Phu Quoc City under three penetration scenarios (2030, 2035, 2040) and by translating these indicators into actionable insights for charging infrastructure and grid planning.

This study focuses on the MCS model to estimate the SCF in a typical EV network in Phu Quoc City. The study analyzes the number of EVs currently in operation in Phu Quoc City and the associated charging behavior of users. The aim is to provide SCF values for different types of EVs and propose strategies to reduce pressure on the grid. Methodologically, the MCS framework also quantifies statistical uncertainty by reporting 95% confidence intervals for all SCF estimates and applying Kolmogorov–Smirnov (K–S) goodness-of-fit tests to verify the internal consistency of the assumed charging distributions. The study's main contribution is to offer practical and long-term planning strategies for adding EVs to smart power grids. The findings also give useful data to help design better charging stations and improve energy demand forecasts. This supports energy planners in choosing better locations for new stations and manage electricity more effectively, supporting grid reliability and efficient capacity planning.

## 2. METHODOLOGY

### *Flowchart of research*

The methodological approach is illustrated in Fig. 1 through a flowchart. It proceeds as follows:

First, the block (1) collects data from Phu Quoc City, including information about population size, transport, and other related factors.

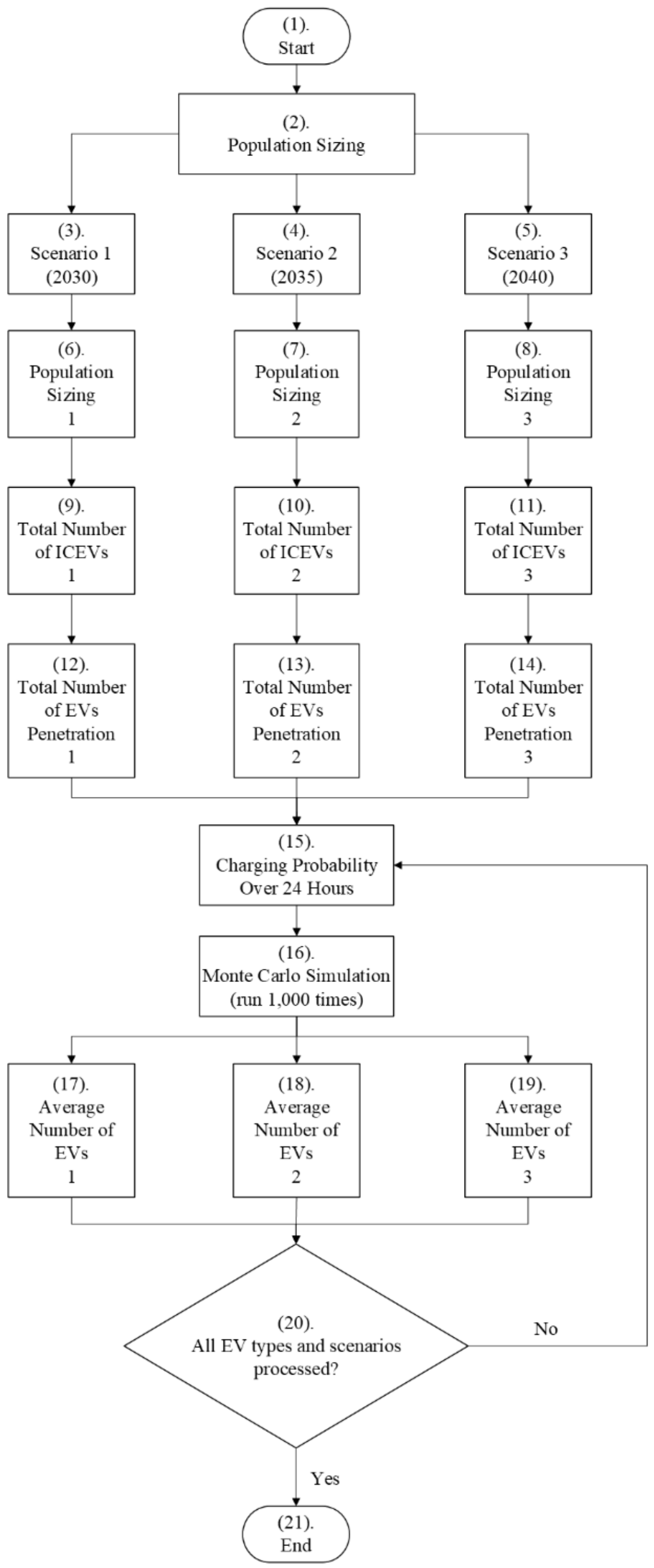

**Fig. 1. Flowchart of the research on the SCF for EVs.**

The block (2) provides the current population data from recent years and uses it to create scenarios for the years 2030, 2035, and 2040; The blocks (3), (4), and (5) represent three different scenarios: a low-level scenario (scenario 1 – 2030), a medium-level scenario (scenario 2 – 2035), and a high-level scenario (scenario 3 – 2040). Each scenario assumes different rates of population growth and transport demand; The blocks (6), (7), and (8) predict the population changes for 2030, 2035, and 2040 based on annual growth rates; The blocks (9), (10), and (11) calculate the total number of ICEVs based on the population and then determine the number of cars, motorbikes, buses, and taxis for each period; The blocks (12), (13), and (14) estimate the number of EVs that will enter the market based on proposed percentages for each vehicle type, the block (15) calculates the probability of EVs charging in 24 hours, based on users charging and travel habits; The block (16) runs a MCS 1,000 times to calculate the average number of EVs charging in 24 hours, shown in blocks (17), (18), and (19); The block (20) then checks whether all EV types and all three scenarios have been processed. Block (20) checks completion across vehicle types and scenarios; Block (21) gives the results and ends the flow. The procedure is repeated for all vehicle types and scenarios. The outputs include the simulated hourly charging counts and the resulting SCF values.

***Mathematical model***

*Linear regression (LR)*

To forecast the future population from recent historical data, an LR model is applied [23]. The model is expressed as $\hat{y} = ax + b$, where $x$ denotes the year index and $\hat{y}$ is the projected population. Detailed derivations are provided in Appendix A.

*Constant Exponential Function (CEF)*

To forecast the future number of vehicles by type, a CEF model is used [24]. The model formulation and notation are provided in Appendix B.

*Binomial Distribution Model (BDM)*

To model the hourly number of charging EVs, a binomial distribution is adopted [25]. The probability mass function and coefficient definition are provided in Appendix C.

*Cumulative Distribution Function (CDF)*

To determine the minimum number of successes $k$ that satisfies a target cumulative probability level, the CDF of the binomial model is used [26]. The CDF expression and the selection criterion are provided in Appendix D. Here, $k$ denotes the minimum number of successes and $\alpha$ is the target confidence level (e.g., 0.95).

*Simultaneous Charging Factor (SCF)*

In this study, SCF is defined as the daily mean proportion of vehicles charging over 24 hours. The SCF of EVs is calculated based on the following formula [27]:

$$SCF = \frac{EV_{avg}}{EV_{total}} \tag{1}$$

where $EV_{total}$ is the total number of EVs (units); $EV_{avg}$ is the average number of vehicles charging within 24 hours (units), calculated using the following formula:

$$EV_{avg} = \frac{1}{24}\sum_{i=1}^{24} X_i \tag{2}$$

where $X_i$ is the number of EVs charging during the hour $i$.

## 3. CALCULATION AND SIMULATION

***Case study***

*Population*

Data on the population of Phu Quoc City for the years 2022, 2023, 2024, and February 2025 have been collected [28]. The simple linear trend population growth formula (A1) is fitted to these data to obtain a neutral 2025–2040 population scenario for the SCF. The calculations are presented in Table 1 and illustrated in Fig. 2. An LR is adopted as a transparent baseline because the available historical population series for the study area is short and the study aims to provide interpretable planning inputs rather than long-term demographic forecasting.

**Table 1. Population forecast to 2040**

| Year | Population (people) |
|---|---|
| Feb-2025 | 150,923 |
| 2025 | 152,217 |
| 2030 | 163,837 |
| 2035 | 175,457 |
| 2040 | 187,077 |

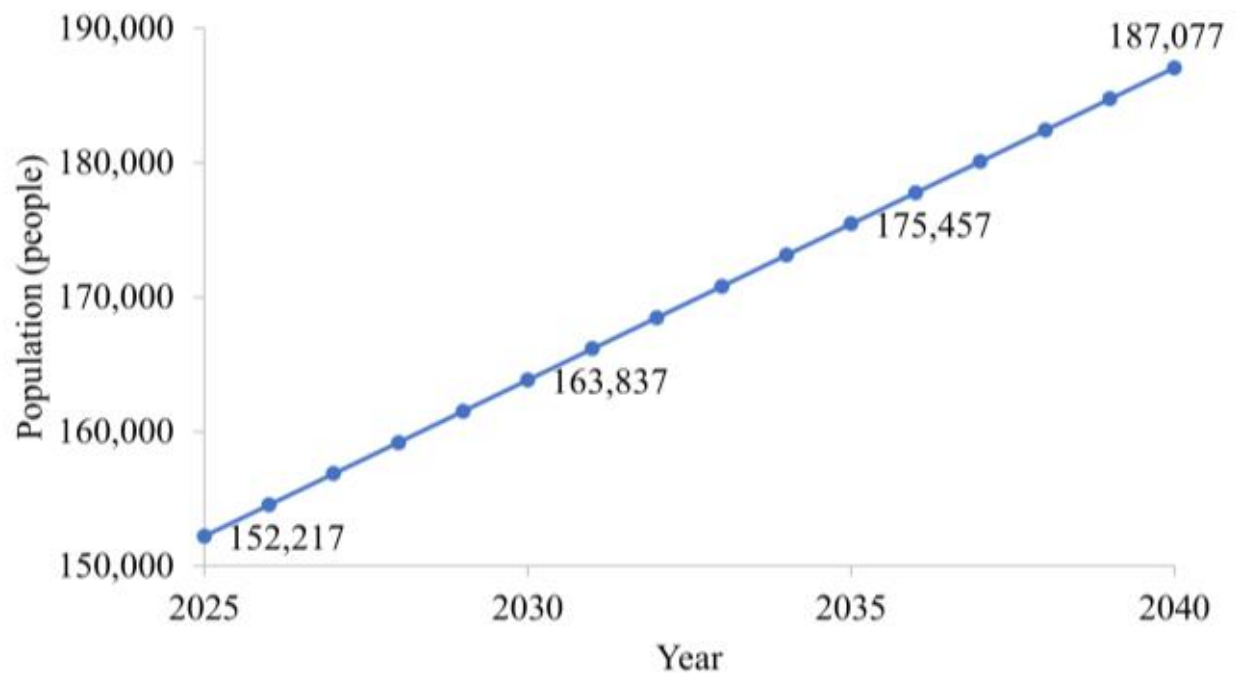

**Fig. 2. Population forecast chart to 2040 (people).**

To test sensitivity to growth assumptions, an exponential population pathway is also evaluated using the CEF model (Appendix B). Calibrated to match the same 2025 baseline and 2040 endpoint as the LR baseline, the exponential pathway yields 163,048 (2030) and 174,650 (2035), compared with 163,837 and 175,457 under LR (differences <0.5%). This indicates that, for the 2030–2040 horizon, the SCF trends reported in this study are not materially affected by the choice between linear and simple non-linear population growth assumptions.

**Table 2. Comparison of LR vs CEF (sensitivity case)**

| Year | LR | CEF | Sensitivity |
|---|---|---|---|
| 2030 | 163,837 | 163,048 | -0.48% |
| 2035 | 175,457 | 174,650 | -0.46% |
| 2040 | 187,077 | 187,077 | 0.00% |

*Internal Combustion Engine Vehicles (ICEVs)*

The current motorcycle penetration rate is 670 per 1,000 people, with an average annual growth rate of 7% [29]. The current car penetration rate is 50 per 1,000 people, with an average annual growth rate of 23.07% [30]. The bus penetration rate is 8 buses per 1,000 people, with an average annual growth rate of 12.13% [30]. The current taxi penetration rate is 90 per 1,000 customers, with an average annual growth rate of 10.25% [31]. The ratio of 5-seat to 7-seat cars and taxis is 50:50, with the assumption that this ratio remains unchanged until 2040. Using formula (B1), the total number of ICEVs forecasted for 2040 is calculated and presented in Table 3, with a visual representation in Fig. 3.

**Table 3. Total number of ICEVs**

| Year | Motorbike (units) | Car (units) | Taxi (units) | Bus (units) |
|---|---|---|---|---|
| 2025 | 101,986 | 7,611 | 13,700 | 1,218 |
| 2030 | 153,960 | 23,128 | 24,019 | 2,323 |
| 2035 | 231,251 | 69,930 | 41,899 | 4,410 |
| 2040 | 345,822 | 210,511 | 72,768 | 8,336 |

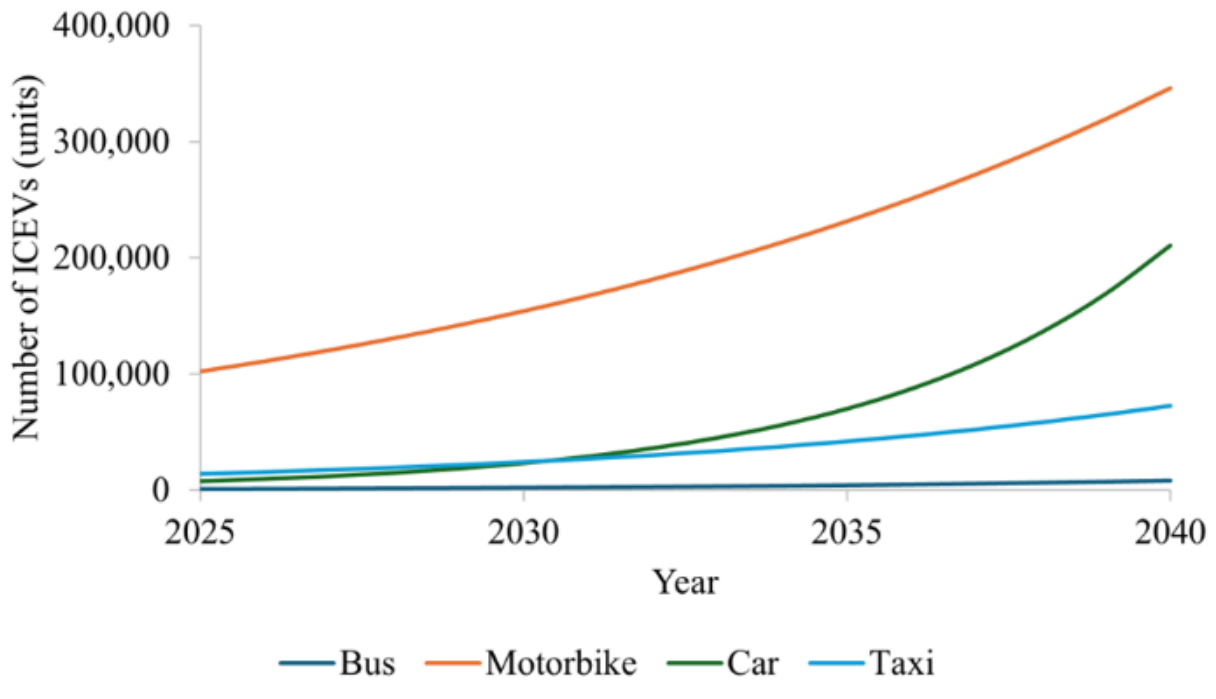

**Fig. 3. ICEVs forecast chart until 2040.**

***Calculation of the SCF of EVs based on MCS***

*Monte Carlo simulation (MCS)*

MCS is a stochastic computer simulation method based on statistical probability theory, which can accurately describe the characteristics of physical objects and experimental processes. It is used to predict and forecast risks and uncertainties [32].

In this paper, MCS is applied based on the random distribution probability of EV charging behavior. Mathematical models (C1), (C2), and (D1) shown in the Appendices are used to derive the probabilities for each type of EV.

*Definition of scenarios*

In accordance with Decision No. 876/QĐ-TTg dated July 22, 2022, in which the Deputy Prime Minister approved the Action Program for the Transition to Green Energy and the Reduction of Carbon Dioxide and Methane Emissions in the Transportation Sector, the EV penetration scenarios used in this study are derived from this directive [33] and are summarised in Table 4.

Based on the total number of ICEVs shown in Table 3 and the scenarios outlined in Table 4, the number of EVs penetrating each scenario is presented in Table 5.

In this study, the numbers of EM, EC, ET and EB in each scenario should be interpreted as stylised scenario pathways rather than precise forecasts. They are derived from a common trend of rising EV penetration, with mode-specific shares chosen to be internally consistent with the expected evolution of the vehicle fleet in Phu Quoc City, rather than from independent extrapolations for each vehicle category, and to maintain transparency about the key underlying modelling assumptions used.

**Table 4. Development of penetration scenarios for different types of EVs**

| Scenario | Electric motorbike | Electric car | Electric taxi | Electric bus |
|---|---|---|---|---|
| Scenario 1 (2030) | 40% | 25% | 50% | 100% |
| Scenario 2 (2035) | 75% | 50% | 100% | 100% |
| Scenario 3 (2040) | 100% | 100% | 100% | 100% |

**Table 5. The total number of EVs penetrating each scenario**

| Scenario | Electric motorbike (units) | Electric car (units) | Electric taxi (units) | Electric bus (units) |
|---|---|---|---|---|
| Scenario 1 (2030) | 61,584 | 5,782 | 12,010 | 2,323 |
| Scenario 2 (2035) | 173,438 | 34,966 | 41,898 | 4,410 |
| Scenario 3 (2040) | 345,822 | 210,511 | 72,768 | 8,336 |

*Charging behavior of different types of EVs*

Electric bus (EB): The operating characteristics of the bus are relatively fixed, with a daily distance of about 70 km. To ensure safe operation, it is typically scheduled to charge twice a day. EBs typically use fast charging at noon and regular charging in the evening after service. Their charging periods are from 09:30 to 16:00 and from 23:00 to 05:00 the next day [34].

Electric motorbike (EM): The operating times of motorbikes are more varied and flexible compared to other types of vehicles. The average daily distance covered by motorbikes is about 21 km. EMs mainly rely on regular charging mode, with charging times around noon and in the evening. The charging periods for EMs are from 11:30 to 13:00 and from 20:30 to 06:30 the next day [35].

Electric car (EC): The operating characteristics of ECs are more variable and flexible compared to buses and taxis. ECs are generally charged once a day, with charging times selected from 8:00 to 12:00, 13:00 to 17:00, and from 19:00 to 06:00 the next day. The average daily driving distance for ECs is about 15 km [36].

Electric taxi (ET): The operating hours of a taxi are from 06:00 to 24:00, with a daily driving distance of about 400 km. ETs require two charging sessions per day, with charging times typically chosen at noon and in the evening. Due to limited rest time, ETs generally opt for fast charging to ensure timely recharging. The charging periods for taxis are from 2:00 to 5:00 and from 11:30 to 14:30 [36].

**Table 6. Probability $p$ for each type of EV within 24 hours**

| Time | EM | EC | ET | EB |
|---|---|---|---|---|
| Hour 1 | 0.50 | 0.50 | 0.20 | 1.00 |
| Hour 2 | 0.50 | 0.50 | 0.30 | 1.00 |
| Hour 3 | 0.50 | 0.50 | 0.40 | 1.00 |
| Hour 4 | 0.50 | 0.40 | 0.50 | 1.00 |
| Hour 5 | 0.40 | 0.40 | 0.50 | 1.00 |
| Hour 6 | 0.30 | 0.10 | 0.20 | 0.01 |
| Hour 7 | 0.20 | 0.05 | 0.05 | 0.01 |
| Hour 8 | 0.10 | 0.40 | 0.05 | 0.01 |
| Hour 9 | 0.10 | 0.40 | 0.05 | 0.20 |
| Hour 10 | 0.10 | 0.40 | 0.10 | 0.20 |
| Hour 11 | 0.50 | 0.40 | 0.20 | 0.10 |
| Hour 12 | 0.50 | 0.20 | 0.30 | 0.20 |
| Hour 13 | 0.50 | 0.10 | 0.50 | 0.10 |
| Hour 14 | 0.40 | 0.20 | 0.50 | 0.20 |
| Hour 15 | 0.30 | 0.40 | 0.40 | 0.10 |
| Hour 16 | 0.20 | 0.40 | 0.10 | 0.10 |
| Hour 17 | 0.10 | 0.05 | 0.05 | 0.01 |
| Hour 18 | 0.10 | 0.05 | 0.05 | 0.01 |
| Hour 19 | 0.10 | 0.10 | 0.10 | 0.05 |
| Hour 20 | 0.40 | 0.30 | 0.20 | 0.05 |
| Hour 21 | 0.40 | 0.30 | 0.20 | 0.05 |
| Hour 22 | 0.40 | 0.40 | 0.30 | 0.05 |
| Hour 23 | 0.40 | 0.50 | 0.30 | 1.00 |
| Hour 24 | 0.50 | 0.50 | 0.40 | 1.00 |

In the MCS, for each scenario and each vehicle type, the number of vehicles charging during an hour $h$ is modeled as a binomial random variable $Binomal(N, p)$, where $N$ is the total number of EVs in that scenario in Table 5 and $p$ given by the corresponding hourly value in Table 6. These probabilities are applied independently across vehicles within each hour and in Monte Carlo runs. Instead of tracking explicit state of charge (SOC) levels, we assume each vehicle charges once or twice per day within the time windows in Fig. 4. For each scenario, 1,000 Monte Carlo iterations are performed, and the simulated hourly charging counts are averaged to obtain the expected number of vehicles charging per hour, which is then used to compute the SCF values via (1)–(2).

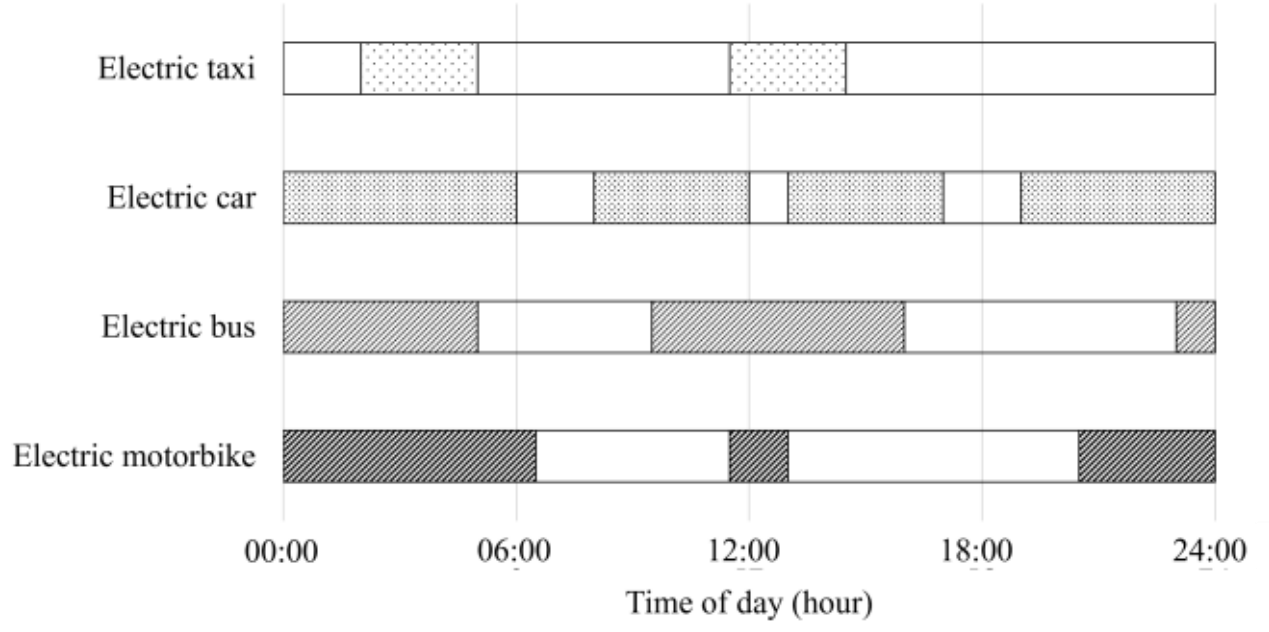

**Fig. 4. EV charging behaviors of each type of EV.**

*Statistical diagnostics and confidence intervals*

To quantify the uncertainty associated with the Monte Carlo estimates, each simulation run is treated as one realisation of the hourly charging profile. For a given scenario and vehicle type, the average number of charging vehicles over 24 hours in run $r$, $E^r_{avg}$ is computed from the simulated hourly counts, and the corresponding SCF value is obtained as $SCF^{(r)} = E^r_{avg} / N_{\mathrm{EV}}$, where $N_{\mathrm{EV}}$ is the total number of electric vehicles in that scenario. The SCF values reported in Tables 8, 10 and 12 are the sample means across the R=1000 runs, and 95% confidence intervals (CIs) are constructed using the standard normal-approximation CI for the sample mean [37] as:

$$SCF_{95\%} = \overline{SCF} \pm 1.96 \frac{s_{SCF}}{\sqrt{R}} \quad (3)$$

where $s_{SCF}$ is the sample standard deviation of $\overline{SCF}$ over the R runs and 1.96 is the 97.5th percentile of the standard normal distribution, corresponding to a two-sided 95% confidence level. Because the number of runs and vehicles is large, the resulting intervals are relatively narrow, but they provide an explicit measure of the residual sampling variability in the SCF estimates.

As a simple goodness-of-fit check, for representative hours spanning high and low demand periods in each vehicle class, we compare the empirical distribution of the simulated hourly charging counts with the theoretical binomial distribution $Bin(N, p)$ assumed in the model. K–S tests applied to these samples do not reject the null hypothesis at the 5% significance level, indicating that the simulated hourly counts are consistent with the underlying binomial assumptions and that the Monte Carlo implementation is internally coherent.

## 4. NUMERICAL RESULTS

### *Low-level scenario – Scenario 1 (2030)*

Using the binomial model (Appendix C), the MCS is run 1,000 times for all EV types. Figs. 5–8 show the hourly charging demand of EMs, ECs, ETs and EBs in Scenario 1 (2030). The profiles are highly peaked: the maximum hourly values are about 31,000 EMs, 3,000 ECs, 6,000 ETs (mainly between 4:00–5:00 and 13:00–14:00) and 2,300 EBs, mostly after 23:00, while the corresponding minimum values are around 6,000, 150 and 270 units, and close to zero for buses.

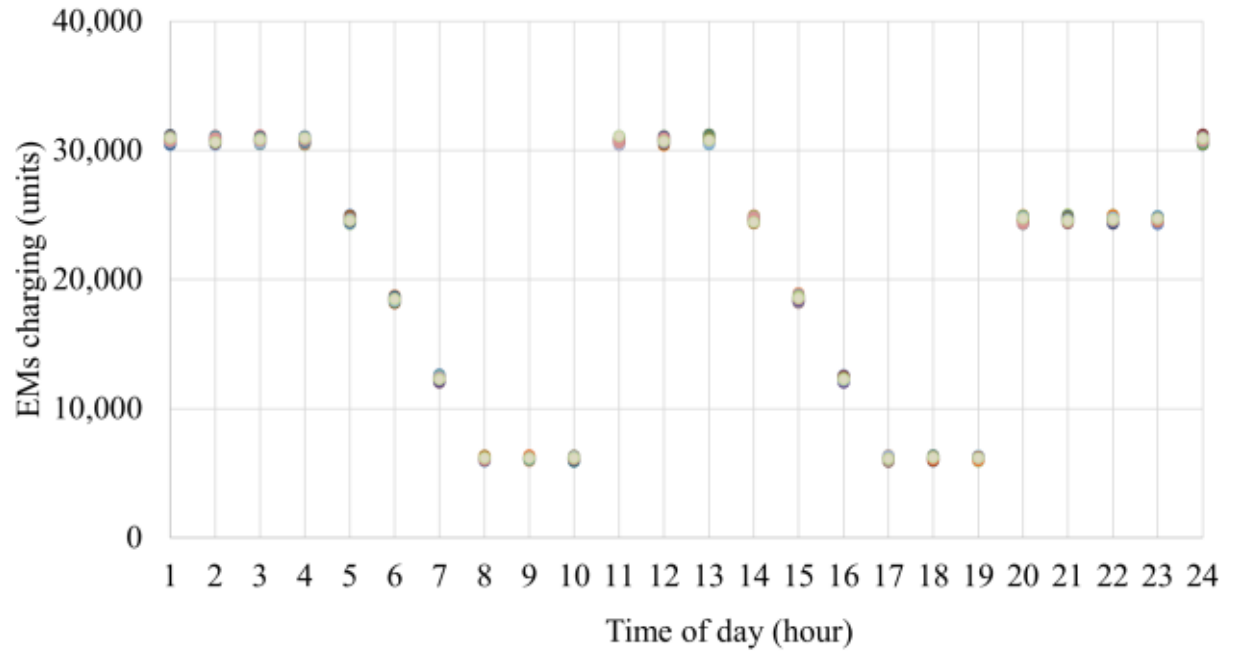

**Fig. 5. EM charging profile over 24 hours – Scenario 1.**

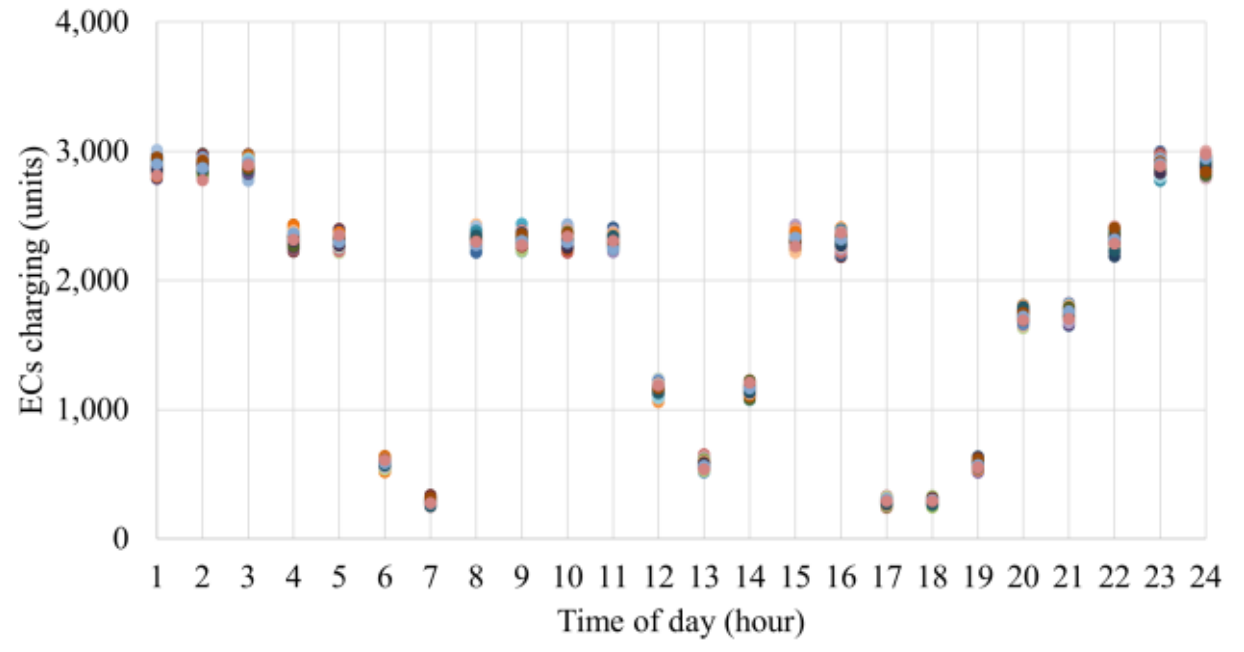

**Fig. 6. EC charging profile over 24 hours – Scenario 1.**

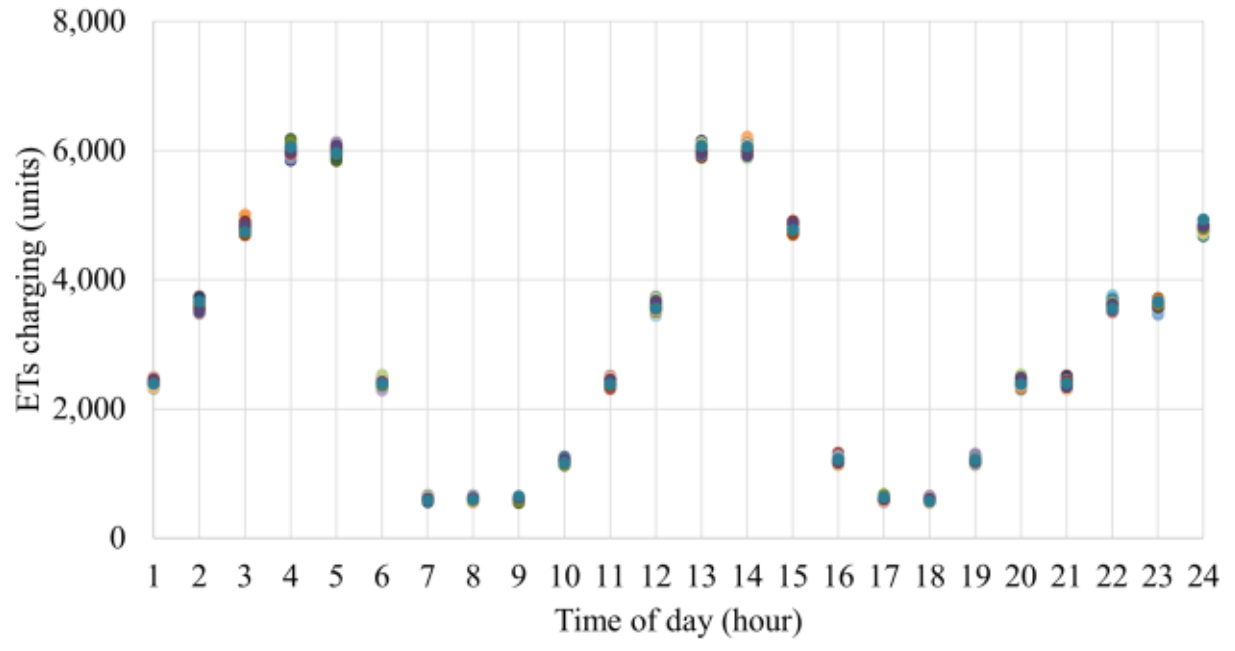

**Fig. 7. ET charging profile over 24 hours – Scenario 1.**

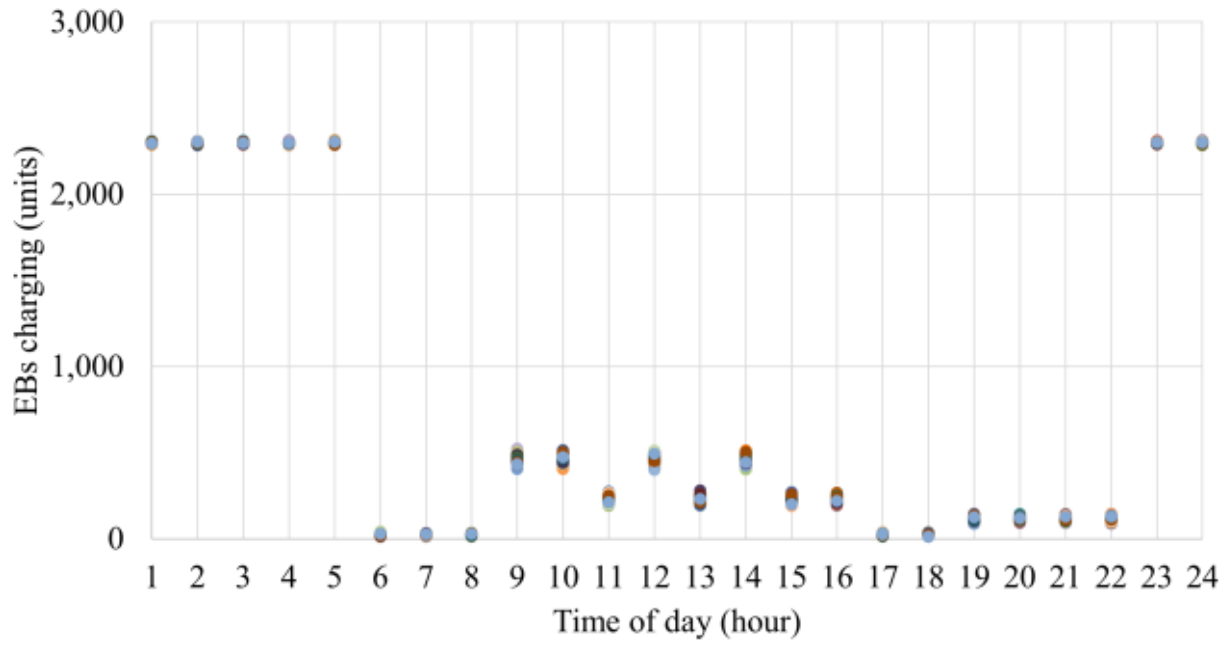

**Fig. 8. EB charging profile over 24 hours – Scenario 1.**

**Table 7. Peak hourly number of EVs charged after 1,000 simulations – Scenario 1**

| EV type | Peak number (units) | Peak hour (hour) |
|---|---|---|
| EM | 31,238 | 12 |
| EC | 3,026 | 2 |
| ET | 6,212 | 13 |
| EB | 2,317 | 2 |

**Table 8. Average number of vehicles charged $E_{avg}$ and SCF with 95% CIs – Scenario 1**

| EV type | $E_{avg}$ (units) | SCF | 95% CI |
|---|---|---|---|
| EM | 20,871 | 0.5181 | [0.51808; 0.51812] |
| EC | 1,925 | 0.7583 | [0.75823; 0.75837] |
| ET | 3,118 | 0.6378 | [0.63775; 0.63785] |
| EB | 843 | 0.7617 | [0.76164; 0.76176] |

Table 7 summarises the peak hourly charging demand and the corresponding peak hour for each vehicle type. Using formulas (1) and (2), the average number of vehicles charged $E_{avg}$ and the SCF for each EV type in the low-level scenario (Scenario 1 – 2030) are calculated and reported in Table 8. The 95% CIs in Table 8 are narrow relative to the differences between vehicle types, indicating that the SCF values are well separated across EMs, ECs, ETs and EBs.

### *Medium-level scenario – Scenario 2 (2035)*

For the medium-level scenario, Figs. 9–12 present the 24-hour charging profiles of EMs, ECs, ETs and EBs, respectively. Compared with Scenario 1, the demand levels increase markedly. The hourly number of EMs requiring charging now exceeds 80,000 at the evening peak and never falls below about 18,000 units.

For ECs, the demand ranges from roughly 800 to 18,000 units, while ETs reach around 21,000 units during two main peaks (4:00–5:00 and 13:00–14:00) and drop to about 1,000 units during off-peak hours. EB charging remains highly concentrated at night, with most buses being charged after 23:00 and a maximum of just over 4,000 units per hour.

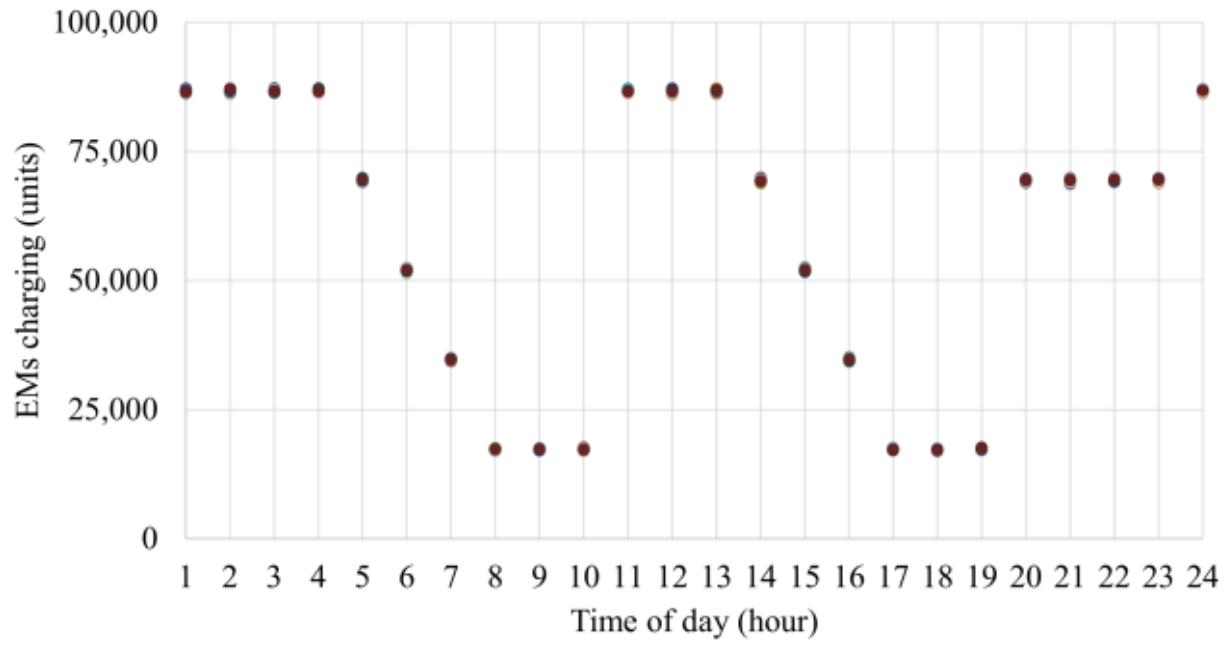

**Fig. 9. EM charging profile over 24 hours – Scenario 2.**

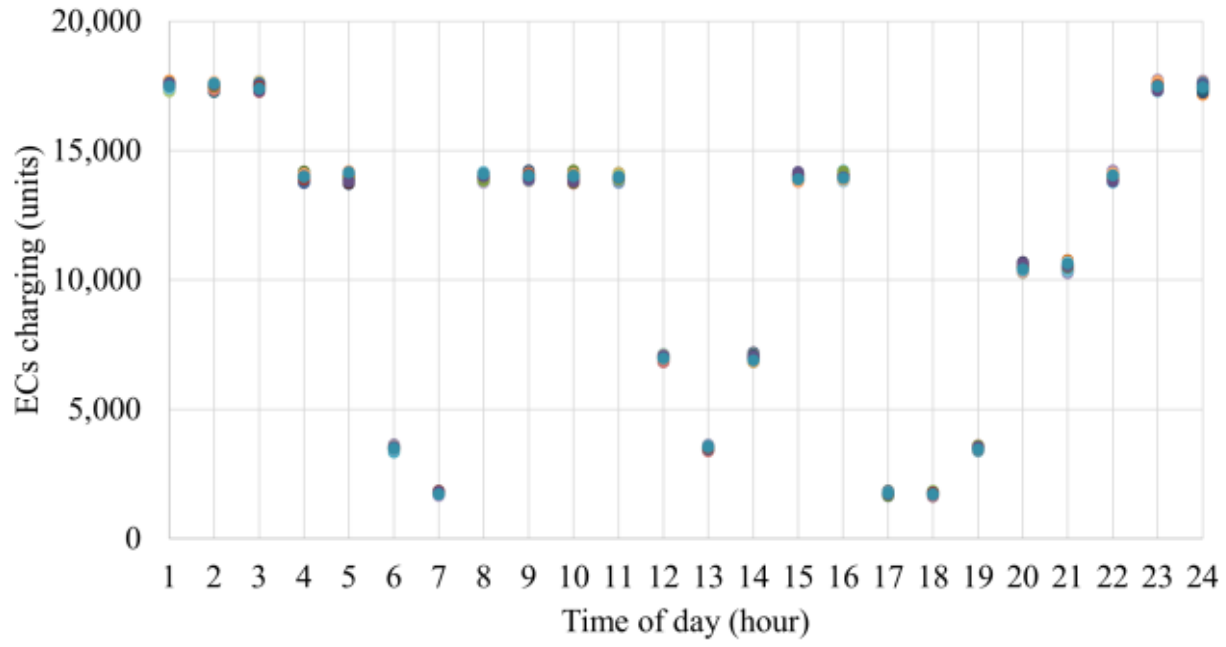

**Fig. 10. EC charging profile over 24 hours – Scenario 2.**

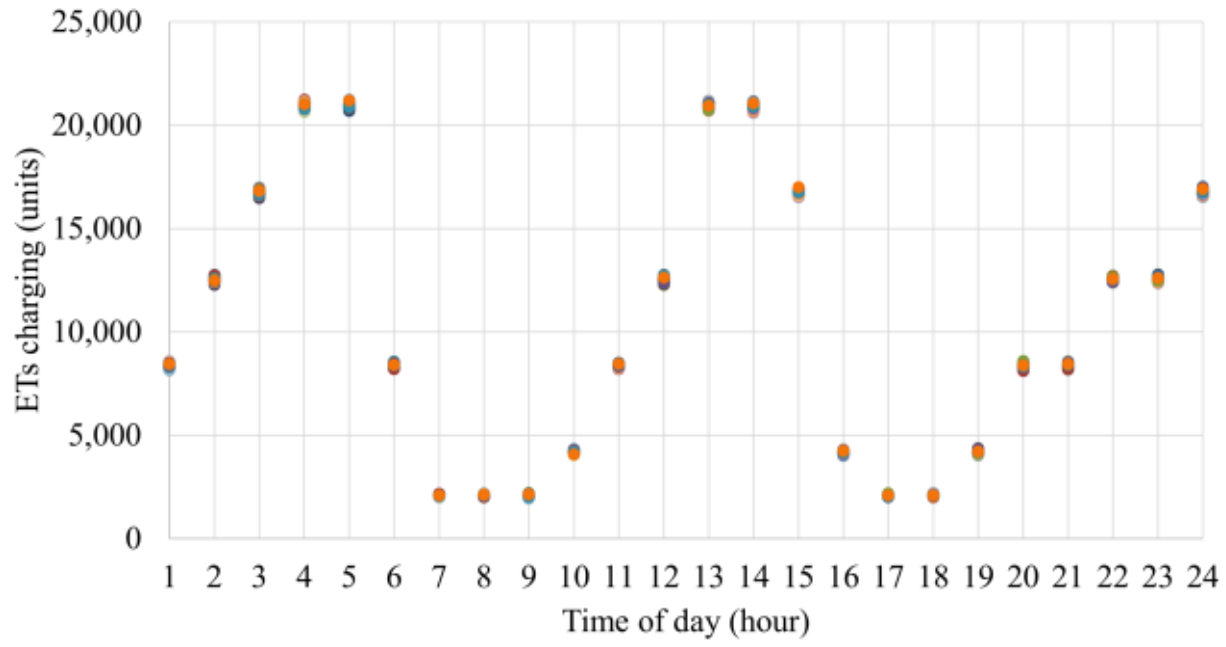

**Fig. 11. ET charging profile over 24 hours – Scenario 2.**

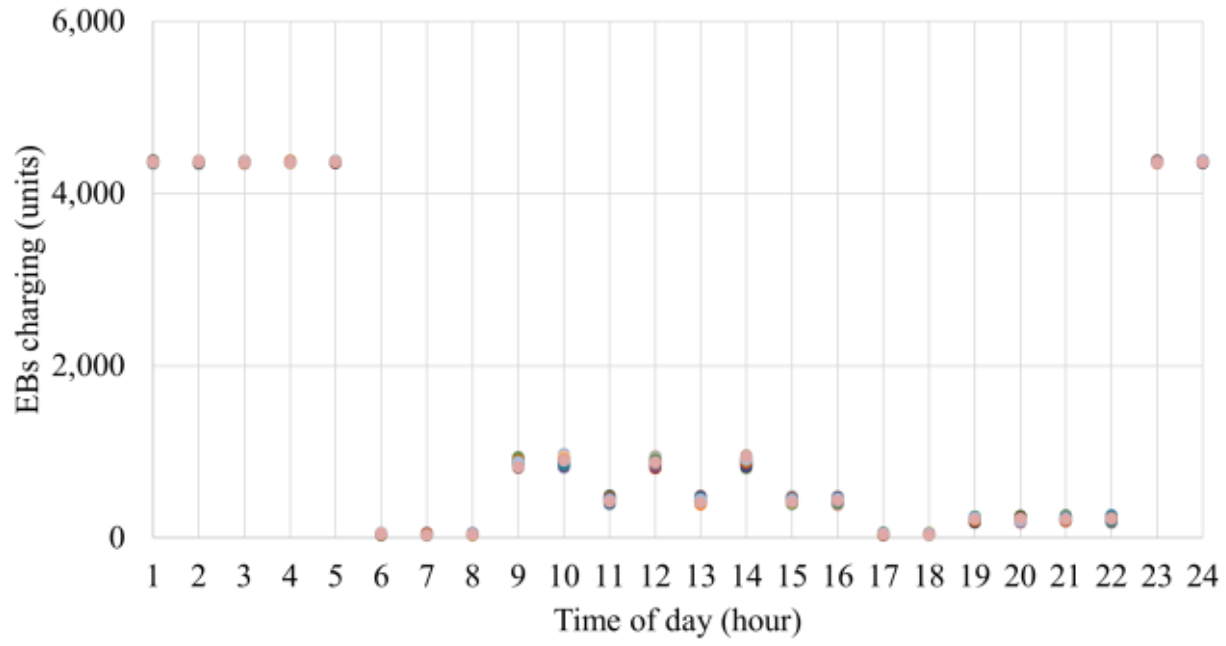

**Fig. 12. EB charging profile over 24 hours – Scenario 2.**

Table 9 summarises the peak hourly demand and the corresponding peak hour for each vehicle type in Scenario 2. Using formulas (1) and (2), the average number of vehicles charged $E_{avg}$ and the SCF are obtained and reported in Table 10. Compared with Scenario 1, the charging demand increases substantially for all vehicle types (e.g., EMs and ECs reach 58,401 and 11,255 units, respectively), while their SCFs decrease to 0.4406 (EM), 0.5514 (EC), 0.4580 (ET) and 0.6246 (EB), indicating that higher EV penetration in 2035 leads to more intensive but slightly more temporally dispersed charging. The CIs in Table 10 confirm that the reductions in SCF between Scenario 1 and Scenario 2 are statistically meaningful rather than an artefact of Monte Carlo noise in the simulation results.

**Table 9. Peak hourly number of EVs charged after 1,000 simulations – Scenario 2**

| EV type | Peak number (units) | Peak hour (hour) |
|---|---|---|
| EM | 87,442 | 13 |
| EC | 17,805 | 2 |
| ET | 21,270 | 14 |
| EB | 4,388 | 3 |

**Table 10. Average number of vehicles charged $E_{avg}$ and SCF with 95% CIs – Scenario 2**

| EV type | $E_{avg}$ (units) | SCF | 95% CI |
|---|---|---|---|
| EM | 58,401 | 0.4406 | [0.44059; 0.44061] |
| EC | 11,255 | 0.5514 | [0.55137; 0.55142] |
| ET | 10,640 | 0.4580 | [0.45797; 0.45802] |
| EB | 1,584 | 0.6246 | [0.62455; 0.62464] |

***High-level scenario – Scenario 3 (2040)***

In the high-level scenario, Figs. 13–16 depict the 24-hour charging profiles of EMs, ECs, ETs and EBs. Compared with Scenarios 1 and 2, demand levels increase sharply across all vehicle types due to higher EV penetration. The hourly number of EMs requiring charging almost reaches 180,000 at the peak and does not fall below about 34,000 units, indicating a high and persistent background load. For ECs, demand varies between roughly 5,200 and 106,000 units, with pronounced morning and evening peaks that overlap with typical residential demand. ETs reach around 37,000 units during two main peaks (4:00–5:00 and 13:00–14:00) and drop to about 2,000 units at off-peak hours, reflecting taxi operating schedules. EB charging remains strongly concentrated at night, with most buses being charged after 23:00, and a maximum of more than 8,000 units per hour. This has important implications for depot supply transformers and upstream feeders.

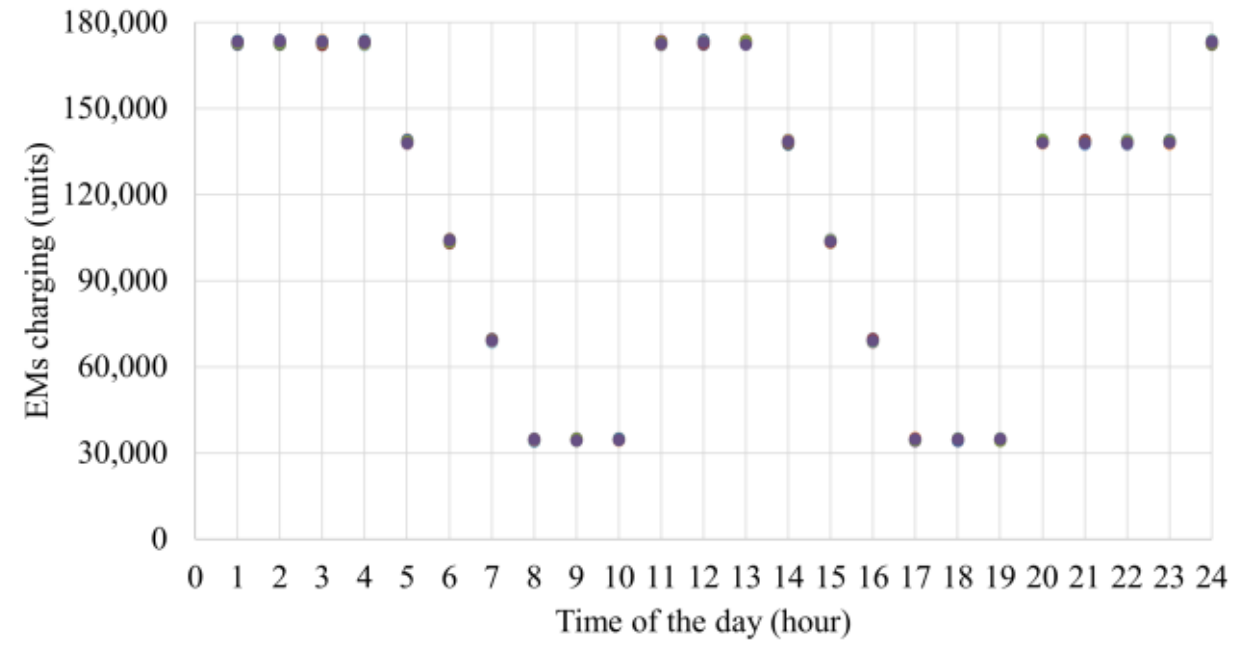

**Fig. 13. EM charging profile over 24 hours – Scenario 3.**

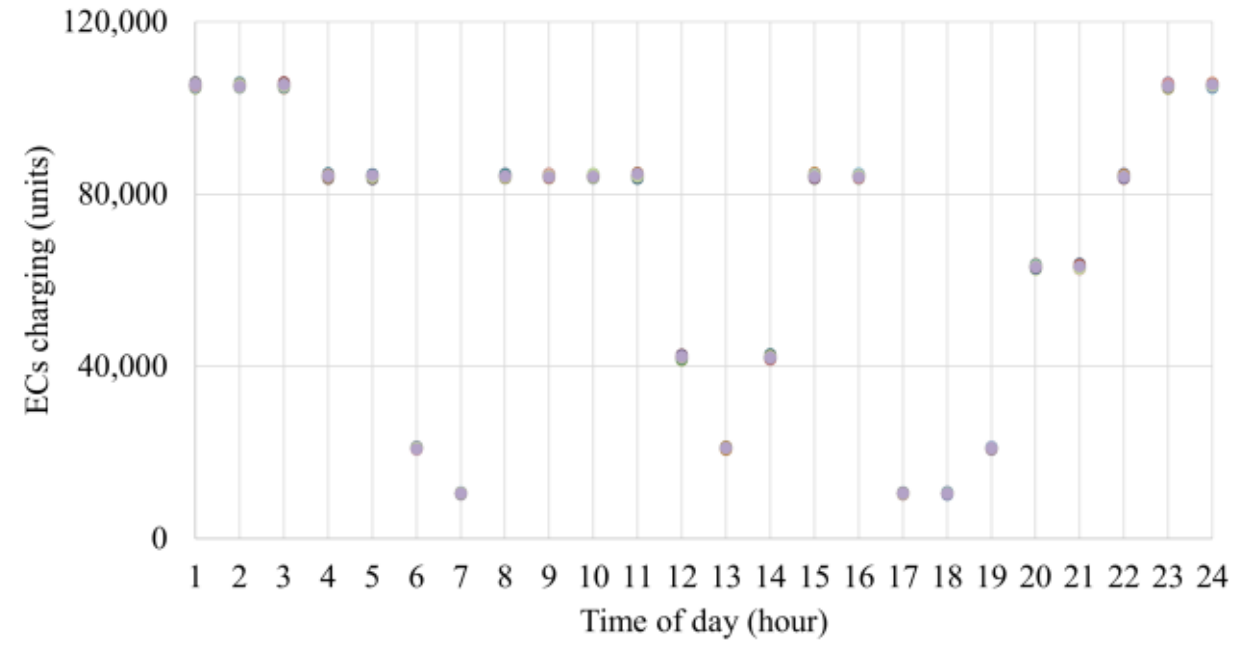

**Fig. 14. EC charging profile over 24 hours – Scenario 3.**

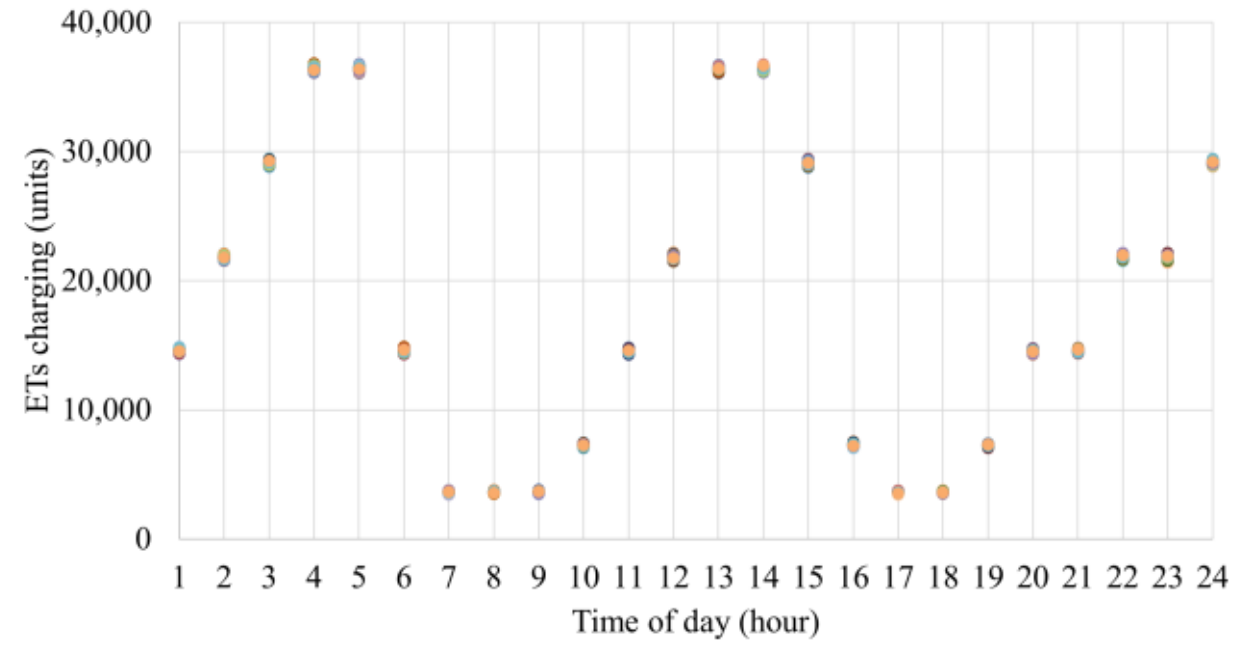

**Fig. 15. ET charging profile over 24 hours – Scenario 3.**

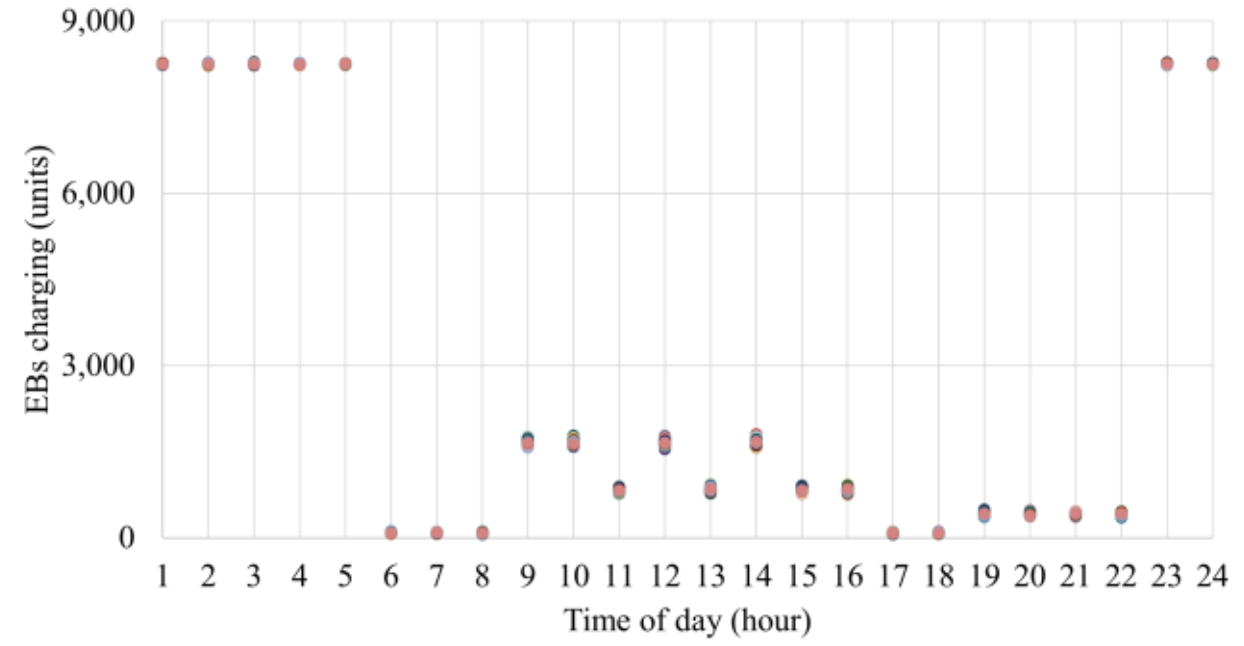

**Fig. 16. EB charging profile over 24 hours – Scenario 3.**

Table 11 summarises the peak hourly charging demand and peak hour for each vehicle type in Scenario 3. Using formulas (1) and (2), the average number of vehicles charged $E_{avg}$ and the SCF are calculated and reported in Table 12. In 2040, $E_{avg}$ reaches 116,128 EMs, 66,853 ECs, 18,378 ETs and 2,972 EBs, while the SCF values decline to 0.4066 (EMs), 0.4528 (ECs), 0.3926 (ETs) and 0.5911 (EBs). This indicates that, under very high EV penetration, charging becomes both more intensive and more temporally dispersed, although bus charging remains concentrated in a narrow night-time window. The 95% confidence intervals in Table 12 are small relative to the SCF differences across scenarios, supporting the robustness of this downward SCF trend as EV penetration increases.

**Table 11. Peak hourly number of EVs charged after 1,000 simulations – Scenario 3**

| EV type | Peak number (units) | Peak hour (hour) |
|---|---|---|
| EM | 174,005 | 11 |
| EC | 106,107 | 24 |
| ET | 36,826 | 13 |
| EB | 8,282 | 3 |

**Table 12. Average number of vehicles charged $E_{avg}$ and SCF with 95% CIs – Scenario 3**

| EV type | $E_{avg}$ (units) | SCF | 95% CI |
|---|---|---|---|
| EM | 116,128 | 0.4066 | [0.40659; 0.40661] |
| EC | 66,853 | 0.4528 | [0.45278; 0.45281] |
| ET | 18,378 | 0.3926 | [0.39258; 0.39261] |
| EB | 2,972 | 0.5911 | [0.59106; 0.59113] |

## 5. DISCUSSION AND CONCLUSION

This study shows that the SCF decreases as EV penetration increases in Phu Quoc City over 2030–2040, implying more temporally dispersed charging demand across all vehicle categories. The EB group provides a clear example, with SCF declining from 0.76 (2030) to 0.59 (2040), while EM, EC, and ET follow similar downward patterns in Fig. 17.

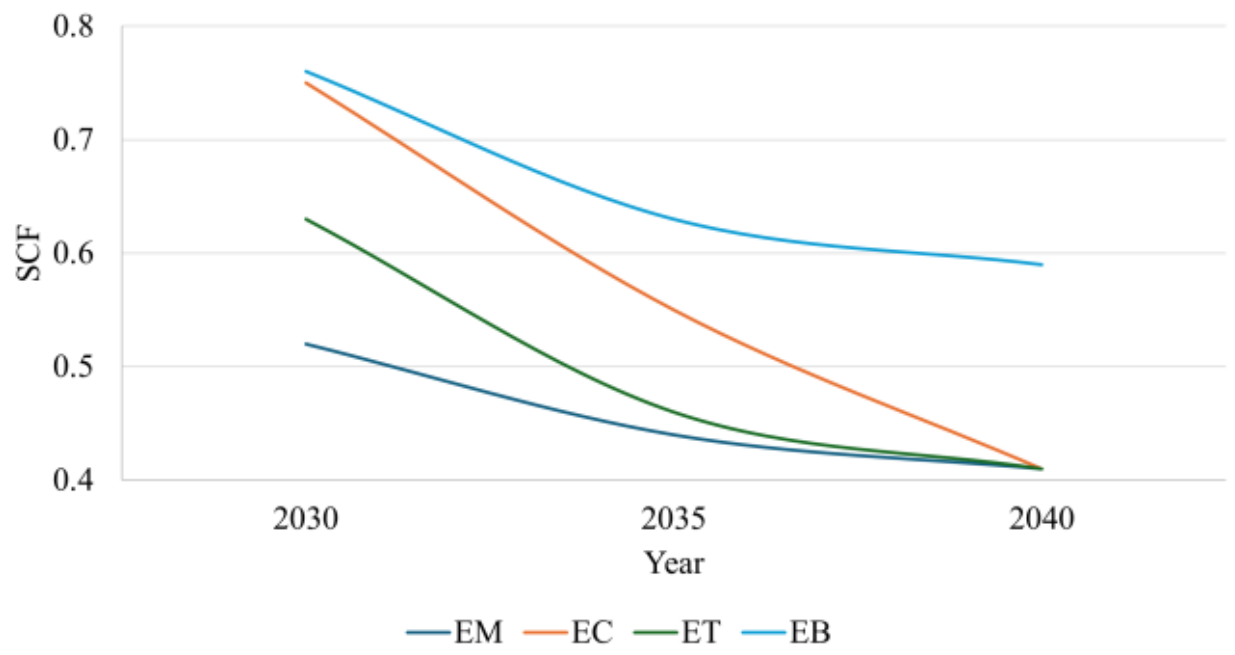

**Fig. 17. SCF curves for three scenarios using the MCS approach**

Across Scenarios 1–3, the 95% CI from 1,000 MCS runs are narrow relative to the SCF differences between years and vehicle groups, indicating that the trends are robust and not driven by Monte Carlo sampling noise. In addition, the statistical diagnostics K–S tests for the representative hours tested do not reject the model assumptions at the 5% level, supporting the internal coherence of the simulation outputs. A non-linear population sensitivity case produces differences below 0.5% relative to the LR baseline, and therefore does not change the main SCF patterns and planning implications.

In line with international findings, EBs and ETs exhibit higher peak simultaneity, whereas ECs and EMs show lower and more diversified charging. This supports adopting vehicle-specific SCF and diversity factors when sizing distribution transformers and feeders, rather than applying a single aggregate factor. Although SCF decreases over time, evening/night peaks remain pronounced for fleet-based charging (especially EBs and ETs) due to tighter operational schedules, while ECs and EMs retain greater flexibility. Therefore, grid-aware planning should combine phased infrastructure expansion with demand management measures such as time of use tariffs, basic smart charging at depots/public stations, and local energy management coordinated with rooftop PV and storage to mitigate peak loading and improve asset utilisation. To provide a simple scenario-based illustration, we consider a peak-shifting factor (PSF) representing the share of flexible charging shifted from peak to off-peak hours via time-of-use pricing or basic smart charging. Assuming daily energy demand is unchanged, a first-order approximation suggests that SCF can be reduced by roughly (1 – PSF) for the shifted portion. For (EB), the 2040 SCF of 0.59 may decrease to approximately 0.53, 0.47, and 0.41 under PSF = 10%, 20%, and 30%, respectively (scenario-based estimate). This illustration is intended for planning insight rather than a full optimisation-based scheduling model. Overall, the proposed SCF framework provides compact planning inputs to support staged deployment of charging hubs, capacity upgrade prioritisation, and more resilient distribution network operation in tourist-intensive cities such as Phu Quoc.

## 6. LIMITATIONS AND FUTURE WORK

This study has several limitations. First, the MCS relies on simplified assumptions about driving distances, operating schedules and charging behavior, using stylised time windows and probabilities that cannot fully capture behavioral diversity or possible smart-charging responses. Second, the data describing current and future EV fleets and population in Phu Quoc City are limited in recency and spatial resolution, and the population path between 2030 and 2040 is represented by a linear interpolation that should be viewed as a neutral baseline rather than a realistic forecast. Third, the absence of anonymised, time-resolved charging records from public and semi-public stations, taxis and buses prevents direct validation of simulated SCF values and loading patterns; while public information on VinFast's network provides useful context, detailed utilisation data are not yet accessible, so the SCF profiles reported here should be interpreted as model-based estimates rather than empirically validated measurements. Finally, although basic statistical diagnostics and CIs are provided, more advanced statistical analyses (e.g., alternative distributional assumptions, bootstrapped intervals) are beyond the present scope.

Future work should prioritise collaboration with VinFast and other charging network operators to obtain anonymised, time-stamped charging data and user surveys, in order to calibrate the MCS and benchmark SCF outputs against observed station utilisation. Integrating IoT-based monitoring and control would allow assessment of smart-charging and demand-response strategies and their impact on SCF and peak loading. The SCF-based demand profiles could also be used to jointly optimise the siting and sizing of PV systems, local storage and fast-charging stations, and to design dynamic pricing and real-time control schemes that respond to grid loading and local renewable output. Extending the framework to multiple cities, high-adoption post-2030 scenarios, and different non-linear demographic and fleet pathways and levels of renewable penetration, combined with structured sensitivity analysis on EV uptake, peak charging periods and battery capacities, would test the robustness of the results and support planning for low-carbon, resilient distribution systems.

## APPENDIX A. Linear Regression (LR)

The least squares estimates for the slope $a$ and intercept $b$ are:

$$a = \frac{\sum_{i=1}^{n}(x_i - \overline{x})(y_i - \overline{y})}{\sum_{i=1}^{n}(x_i - \overline{x})^2}; \qquad b = \overline{y} - a\overline{x} \tag{A1}$$

Hence, the fitted value at $x$ is:

$$\hat{y} = ax + b \tag{A2}$$

where $x_i$ and $y_i$ are observed data; $\overline{x}$ and $\overline{y}$ are sample means; and $n$ is the number of observations.

## APPENDIX B. Constant Exponential Function (CEF)

The constant exponential growth model used in this study is expressed as:

$$F_n = F_0(1+x)^{(n-t_0)} \tag{B1}$$

where $F_0$ is the baseline value at year $t_0$ ; $F_n$ is the projected number of vehicles at year $n$ ; $x$ is the annual growth rate; and $(n-t_0)$ is the time interval in years.

## APPENDIX C. Binomial Distribution Model (BDM)

The hourly number of charging EVs is modeled as a binomial random variable. Its probability mass function is:

$$P(X = k) = \binom{n}{k} p^k.(1-p)^{n-k} \tag{C1}$$

where $X$ is the number of charging EVs in a given hour, $n$ is the number of trials, $p$ is the charging probability, and $k$ is the number of successes. The binomial coefficient is defined as:

$$\binom{n}{k} = \frac{n!}{k!(n-k)!} \tag{C2}$$

## APPENDIX D. Cumulative Distribution Function (CDF)

For a binomial random variable with parameters $n$ and $p$ the CDF is:

$$F(k;n,p) = \sum_{i=0}^{k} \binom{n}{i} p^i (1-p)^{n-i} \tag{D1}$$

where $F(k;n,p)$ denotes the probability of obtaining at most $k$ successes in $n$ trials. For a target probability level $\alpha$ , the minimum $k$ is selected as:

$$k = \min\{k : F(k;n,p) \ge \alpha\} \tag{D2}$$